\documentclass[preprint2]{aastex}

\newcommand{\kms}{km\,s$^{-1}$}
\def\hi{H\,{\sc i}}

\received{2003 November 15}
\begin{document}
\title{High-resolution imaging of the OH megamaser emission in IRAS\,12032+1707 and IRAS\,14070+0525}
\author{Y. M. Pihlstr\"om \altaffilmark{1}} \affil{National Radio
Astronomy Observatory, P.O. Box O, Socorro, NM 87801}
\email{ypihlstr@nrao.edu}

\author{W. A. Baan\altaffilmark{2}} \affil{ASTRON, Oude
Hoogeveensdijk 4, 7791 PD Dwingeloo, The Netherlands}

\author{J. Darling\altaffilmark{3}} \affil{Carnegie Observatories, 813
Santa Barbara Street, Pasadena, CA 91101}

\and

\author{H. -R. Kl\"ockner\altaffilmark{2,4}} \affil{Kapteyn Institute,
University of Groningen, P. O. Box 800, 9700 AV Groningen, The
Netherlands}

\begin{abstract}
  We present results from VLBA observations of the 1667 and 1665~MHz
  OH lines in two OH megamaser galaxies IRAS\,12032+1707 and
  IRAS\,14070+0525. For IRAS\,12032+1707 we also present Arecibo \hi\,
  absorption data. Almost all OH emission previously detected by
  single dish observations has been recovered in IRAS\,12032+1707 and
  is found on a compact scale of $<$100 pc. The emission shows an
  ordered velocity field that is consistent with a single disk
  However, the data is also consistent with a scenario including two
  physically different gas components. We explore this possibility, in
  which the two strongest and most blueshifted maser features are
  identified as the tangent points of a circumnuclear torus.  The
  redshifted maser features would be extended in a direction
  perpendicular to the torus, which is in the direction of the merger
  companion. Thus, redshifted emission could be associated with an
  inflow triggered by a tidal interaction. \hi\ absorption covers the
  velocities of the redshifted maser emission, suggesting a common
  origin. In the second source, IRAS\,14070+0525, a large fraction of
  the OH emission is resolved out with the VLBA.  We find no
  significant evidence of an ordered velocity field in this source,
  but these results are inconclusive due to a very low signal-to-noise
  ratio.
\end{abstract}

\keywords{galaxies: individual (IRAS 14070+0525, IRAS 12032+1707) --- galaxies: starburst --- masers --- galaxies: nuclei --- radio lines: galaxies --- techniques: interferometric}

\section{Introduction}

Since the discovery of OH megamaser emission in Arp\,220
\citep{baan82}, the detection of $\sim$95 OH megamaser have been
reported \citep{baan98,darling02,martin88,staveley87}. These single
dish studies show that the shapes of the OH spectral lines are often
complex, and cover velocities in the range 10\kms\ up to a few 1000
\kms in each source.

Subsequent VLBI investigations of a few nearby OH megamaser galaxies,
such as III\,Zw\,35 and Mrk\,231, have demonstrated that the bulk of
the OH maser emission arises in circumnuclear disks or tori
\citep{pihlstrom01,kloeckner03}.  However, there are also indications
that a single disk component cannot account for {\it all} maser
emission, e.g.\ in Mrk\,273; \citep{kloeckner04}. Early single dish
studies suggested the presence of blueshifted wings in several OH
megamaser spectra \citep{baanetal89}. These have been interpreted as
outflows, which possibly could provide an explanation for the emission
that does not fit comfortably within a disk model.

The existence of outflows as well as inflows is not surprising, given
that OH megamaser galaxies are exclusively associated with
(Ultra)Luminous Infra-Red Galaxies ([U]LIRGs). (U)LIRGs probably
represent short periods of nuclear starburst activity triggered by
merger events. Such mergers will rapidly transport gas to the central
regions, causing gas to fall inwards towards the nuclei. The combined
effects of supernova explosions and stellar winds generated in this
nuclear starburst can entrain the interstellar medium in outflows with
velocities of $100-1000$ \kms\ \citep{heckman90, alton99}.

It is also possible that OH maser emission occurs in both of the
merging nuclei, as has been seen in Arp\,220 \citep{diamond89}.  With
a slight offset in systemic velocity between the two merging nuclei,
the combined spectrum observed with a single dish telescope would be
relatively broad. However, for the very broadest OH megamaser lines
(exceeding 1000 \kms\,), it is hard to interpret the lines either as
originating from a single disk component or from a pair of nuclei.  A
combination of orbital mechanics, possible disk rotation, molecular
inflow and outflow, and the velocity separation of the two main lines
at 1667~MHz and 1665~MHz is likely to make up the velocity range in
such OH megamaser spectra.

Earlier global VLBI experiments have reported the presence of 100
\kms\, broad OH maser lines on parsec scales \citep{diamond99}. In
this paper we will concentrate on investigating the cause of OH
megamaser lines with widths exceeding 1000 \kms. We report on VLBA
observations of two OH megamaser galaxies, IRAS\,14070+0525 and
IRAS\,12032+1707, that have full width zero intensity velocities of
$1500-2000$\,\kms. These objects are additionally interesting because
of their high redshifts ($z=0.217$ and $z=0.265$).  Furthermore, their
high OH luminosities of $L_{\rm OH}=1.3\times 10^4L_{\odot}$ and
$L_{\rm OH}=1.2\times 10^4L_{\odot}$ respectively make the term
'gigamaser' suitable, and as such these galaxies are two of the most
powerful OH megamaser galaxies known. The main aim of the current
observations was to determine whether the broad lines are associated
with inflows, outflows, rotating structures or violent kinematics.
This paper also presents Arecibo \hi\, absorption data for one of the
sources, IRAS\,12032+1707.


\section{Observations}\label{observations}
\subsection{VLBA 1667/1665 MHz OH maser}\label{vlbaobs}
IRAS\,12032+1707 was observed in phase-referencing mode with the VLBA
for 12 hours on 2002 July 8. The redshift of IRAS\,12032+1707
($z=0.217$) shifted the 1667.359\,MHz line to 1369\,MHz, that was used
as the center of the observing band. To cover the complete OH emission
velocity range of IRAS\,12032+1707 ($\sim$2000 \kms), a bandwidth of
16\,MHz per left and right hand polarization was used.  All telescopes
were available at the time of the observations, and only minor periods
of time required flagging due to radio frequency interference (RFI).

IRAS\,14070+0525 was observed with a similar setup in June 9, 2002.
The 16\,MHz IF pair was centered on the redshifted ($z=0.265$)
frequency of 1318\,MHz. Due to RFI the KP telescope showed extreme and
highly variable system temperatures, and to a large extent data had to
be flagged for this telescope. Furthermore, the LA antenna was broken
and so did not participate in these observations. At the observing
frequency of 1318\,MHz, several RFI spikes could be seen in the
autocorrelation spectra. Since the frequencies at which these spikes
occurred differed between the sites, they did not affect the cross
correlation spectra for the most part. Badly affected time ranges were
removed on `by baseline' basis.

The data were correlated at the VLBA correlator in Socorro using the
maximum 512 channels available. Data reduction was performed within
AIPS. Both datasets were amplitude calibrated using the system
temperature measurements, and fringe-fitting was performed on the
phase-reference sources since the low signal to noise ratio (SNR) for
the weak maser features prevented self-calibration on any individual
maser peak. For IRAS\,12032+1707 the data were imaged using robust
weighting, with a beam size $11\times6$ mas ($38\times21$
pc)\footnote{Using $H_0$=$71$ \kms\,Mpc$^{-1}$, $\Omega_{\rm M}=0.27$
  and $\Omega_{\rm vac}=0.73$ in a flat universe, the linear scale is
  3.482 pc\,mas$^{-1}$ at $z$=0.217 and 4.046 pc\,mas$^{-1}$ at
  $z$=0.265.}. IRAS\,14070+0525 was imaged with natural weighting with
a beam size of $13\times6$ mas ($53\times24$ pc)$^1$.  Before the
cubes were analyzed, data were smoothed in frequency in order to
further improve the SNR, to a channel increment of 34 \kms\ for
IRAS\,12032+1707 (1$\sigma$ noise level per channel of 0.6 mJy
beam$^{-1}$) and 82 \kms\ for IRAS\,14070+0525 1$\sigma$ noise level
per channel of 0.6 mJy beam$^{-1}$).

\subsection{Arecibo 21cm \hi\, absorption}\label{areciboobs}
To compare the velocity distribution of the OH emission with any
possible \hi\, absorption, data was taken at Arecibo on June 26, 2003.
Observations were performed in position-switched mode with the L-wide
receiver, covering a bandwidth of 12.5~MHz with 1024 spectral
channels. Hanning smoothing was applied, and left and right
polarizations were averaged to optimize the sensitivity. The total
on-source integration time was 36 minutes, yielding an rms noise of
0.48 mJy.


\section{Results}\label{results}
In order to determine the physical origin of the broad lines, we are
interested in both the spatial distribution and the velocity fields of
our sources. We present results of the data analysis for each source
below.

\subsection{IRAS\,12032+1707 - OH emission}\label{ohspatial}

\begin{figure}[tb]
  \epsscale{1} 
\plotone{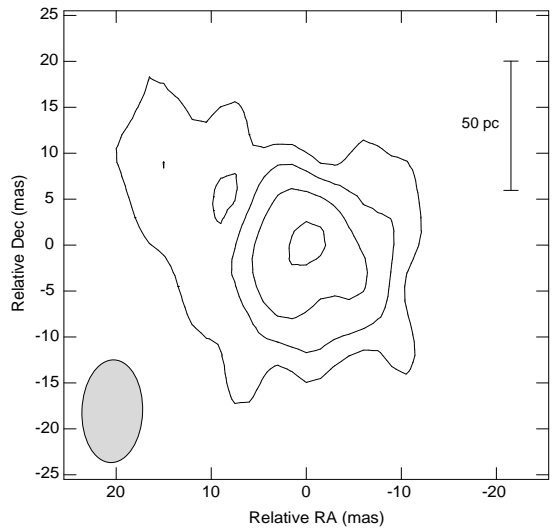}
\caption{The OH maser emission distribution in IRAS\,12032+1707, averaged over all channels with line emission. Contour levels are plotted starting at a 3$\sigma$ level of 0.9 mJy/beam and then increasing by a factor of $\sqrt 2$. The beam size indicated in the lower left corner is $11.3\times 6.46$ mas. Given the SNR in our data, we do not consider the extension to the north-east as significant.\label{ir12_average}}
\end{figure}

Figure \ref{ir12_average} shows the spatial extent of OH megamaser
emission in IRAS\,12032+1707, averaged over all channels of signal
above the $3\sigma$ level. In addition to a central structure confined
within a region of 25$\times$25 mas ($\sim87\times$87 pc), this map
shows an extension of the source to the north-east. The maser line was
to weak to perform self-calibration, and on larger scales than that
shown in Fig.\ \ref{ir12_average} we can see some weaker features in
the map that are likely to be the results of calibration errors. The
north-east extension is only seen at contours close to 3$\sigma$, and
it is uncertain whether this feature is significant.  Mapping the line
emission at a few different resolutions, and using different weighting
schemes show ambiguous results for this feature while the central
component remains the same. Therefore we will not consider the
north-east extension to be significant in the discussion in this
paper.

We conclude that the OH maser emission is centered at RA 12:05:47.7225
and Dec 16:51:08.266 (J2000). This coincides, within the $\sim
0.6\arcsec$ VLA positional errorbars, with the position given for the
continuum in the NRAO VLA Sky Survey (NVSS, \citet{condon98}).
Near-infrared (NIR) and optical imaging have identified two nuclei in
this source \citep{veilleux02} with a projected separation of
3.8\arcsec (13.2 kpc) in an approximately north-south direction. From
the NIR and the optical positions, we can thus associate all OH
megamaser emission with the northern nucleus.

\begin{figure}[tb]
\epsscale{1} 
\plotone{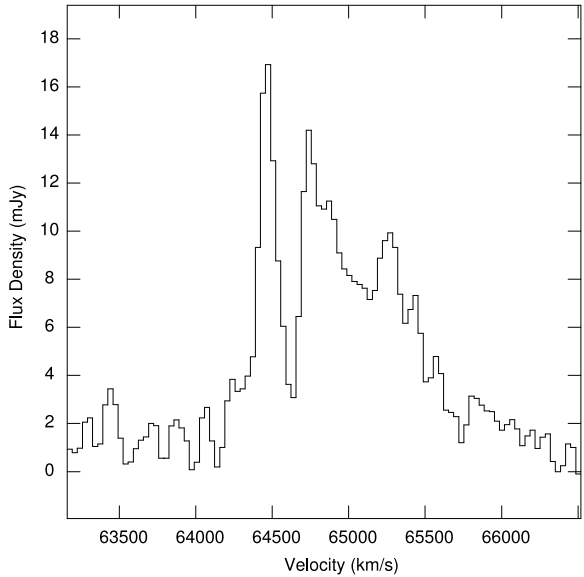}
\caption{The total intensity spectrum of IRAS12032+1707. The flux density levels agree within the error bars to the flux density detected in earlier single dish spectrum. This indicates that we recover all emission in our VLBA data. \label{ir12_totspec}}
\end{figure}

In Fig.\ \ref{ir12_totspec} we plot the resulting spectrum including
the emission in every pixel with emission above $3\sigma$. The
velocity scale is referenced to the 1667~MHz line, and the optical
systemic velocity is 65055 \kms\ \citep{kim98}. This can be compared
to the single dish spectrum published by \citet{darling01}.  Within
the noise, we recover all of the single dish flux density in our VLBA
data. There is a weak, redshifted wing at velocities above 65700
\kms\, in the spectrum by \citet{darling01} that is not clearly
identified in our data. The 3$\sigma$ rms noise level in our spectrum
is at 1.8 mJy/beam, which is at the same level as the emission in the
Arecibo spectrum. Thus, we cannot determine whether the red wing is
resolved out in the VLBA data, or whether it is hard to see because it
is not discernible from the noise.

\begin{figure}[tb]
  \epsscale{1} 
\plotone{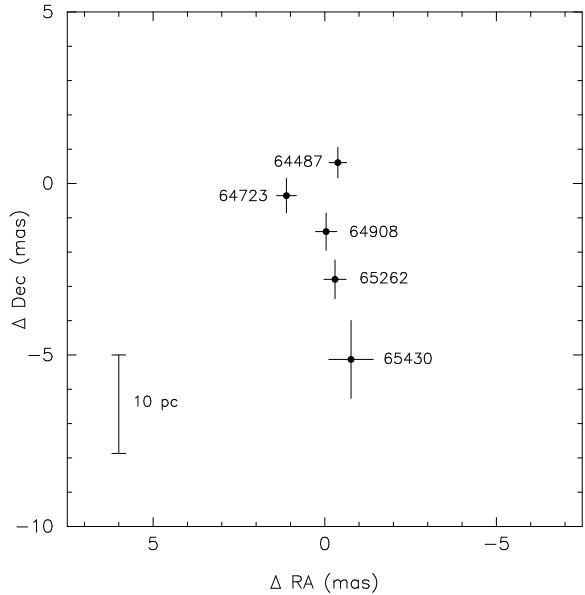}
\caption{The spatial positions of the maser emission in IRAS\,12032+1707. Each point represents a peak in the spectrum, and is plotted with $1\sigma$ positional error bars.\label{ir12_radecpos}}
\end{figure}

Since the SNR is limited in our data, we perform additional spectral
averaging to extract velocity information from the cube. At least five
peaks in our total intensity spectrum can readily be identified in the
single dish spectrum. By averaging a range of six channels around each
peak, we constructed five contour maps.  2-dimensional Gaussian
fitting was then performed on each map to derive their centroid
position. The positional error bars were calculated using the
expression $\sigma_{x,y}=\theta_{x,y}/(2\,SNR)$ where $\theta_{x,y}$
is the angular resolution in each direction, and the SNR is the SNR in
each map. The result of the fitting is displayed in Fig.\ 
\ref{ir12_radecpos}. Clearly there exists an ordered velocity field,
with a major gradient seen in an approximately north-south direction.

The broad velocity extent of the OH emission implies the possibility
of blending between hyperfine transitions at 1667 and 1665~MHz. In the
heliocentric velocity frame these hyperfine lines are expected to have
a velocity separation of 428\,\kms. It could therefore be possible
that the most blueshifted peak at 64487 \kms\ has its 1665~MHz
counterpart in the peak at 64908 \kms. We note however that for the
only cases in which the 1665~MHz emission has been detected in an VLBI
experiment \citep{pihlstrom01, kloeckner04}, the 1665~MHz emission
closely follows the 1667~MHz distribution. Since the two peaks in our
experiment have very different spatial positions, we will assume that
the emission line is dominated by 1667~MHz emission. This is known to
be the case for all OH megamasers with well separated hyperfine
transitions.

\subsection{IRAS\,12032+1707 - continuum}\label{cont}
From the NVSS IRAS\,12032+1707 is expected to have a correlated 1.4
GHz continuum flux density of 29 mJy \citep{condon98}. Due to the
broad line width of the OH maser emission, there was only a limited
part of the observed band that contained line free channels. 20\% of
the total passband was selected and averaged in an attempt to search
for the continuum emission. Using the same imaging parameters as
described for the OH emission (Sect.\ \ref{vlbaobs}) no significant
continuum emission was detected at a $3\sigma$ level of 0.3 mJy/beam.
Given the beam size this implies that the continuum becomes resolved
on scales less than around 75 mas (260 pc).

We note that using natural weighting, in combination with a heavy
tapering of the data, yields a tentative detection of $\sim$1 mJy
coincident with the location of OH emission.  However, given that no
self-calibration can be performed, the significance of this continuum
emission is questionable. In future, more sensitive observations are
required to address this issue.

\subsection{IRAS\,12032+1707 - \hi}\label{HIres}
\begin{figure}[t!]
\epsscale{0.7}
\rotatebox{-90}{
\plotone{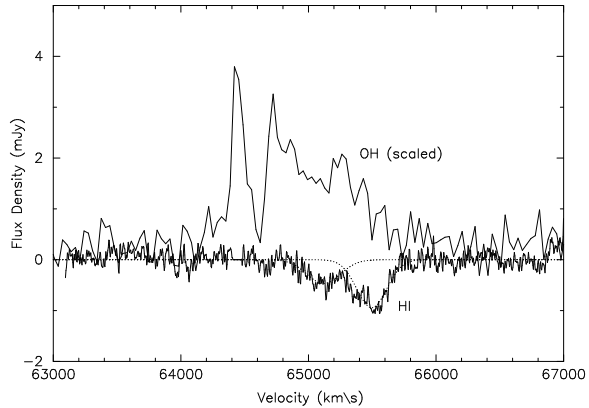}\vspace*{0.4cm}}
\caption{\hi\, absorption spectrum (thick line) showing that the
  \hi\, absorption in IRAS\,12032+1707 corresponds to the reddest
  parts of the VLBA OH maser emission spectrum (thin line). The
  amplitude of the OH spectrum has been divided by a factor of 5, to
  allow a better comparison. In the \hi\, spectrum a region around
  64450 \kms\ has been blanked due to RFI. \label{arecibospectrum}}
\end{figure}

The Arecibo \hi\, absorption spectrum is shown in Fig.\ 
\ref{arecibospectrum}. The \hi\, absorption can be fitted by two
Gaussian components with centroid velocities of 65119 and 65496 \kms\ 
and full width half maxima of 186 and 166 \kms, respectively. Overlaid
on the \hi\, spectrum is the VLBA OH maser spectrum (scaled down by a
factor of 5 in amplitude), allowing a comparison between the \hi\, and
OH velocities. We conclude that the bulk of the \hi\, absorption
coincides in velocity with the reddest velocity components of the VLBA
OH maser emission. The implied \hi\, column densities can be estimated
to $N_{\rm HI}>5.5\times 10^{20} (\frac{T_{\rm sp}}{100 K}) f^{-1}$
cm$^{-2}$, where $T_{\rm sp}$ is the spin temperature and $f$ is the
filling factor.

\subsection{IRAS\,14070+0525 OH emission}\label{res14}

\begin{figure}[t]
\epsscale{1}
\plotone{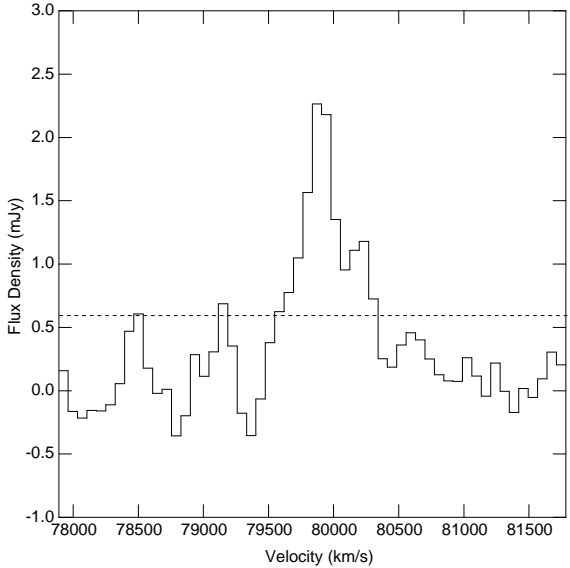}
\caption{Total integrated OH maser emission in IRAS\,14070+0525. The
  flux density is only a fraction of the flux density detected using a
  single dish. The dashed line across the spectrum shows the 1$\sigma$
  noise limit.\label{ir14_totspec}}
\end{figure}

Figure \ref{ir14_totspec} shows the total integrated OH flux density
that is recovered in our VLBA observations. As with IRAS\,12032+1707,
all velocity scales are referenced to the 1667~MHz line. The systemic
velocity of 79445$\pm70$ \kms\ \citep{kim98} is slightly offset from
the velocity components detected in our VLBA data.  We can define two
peaks in the spectrum, with corresponding velocity centroids of
$\sim$79900 \kms\, and 80250 \kms.  Given the low spectral resolution
of our VLBA spectrum (82\,\kms), these peaks coincide with peaks seen
in the single dish spectrum at velocities of 79950\,\kms\, and
80200\,\kms\ \citep{baan92}. However, several distinct single dish
peaks are missing from our VLBA spectrum, and we can only account for
a minor part of the flux density observed by single dish ($<10\%$).
For example, the single dish spectrum displays a prominent feature at
around 79000\,\kms\ that we do not detect. This implies that a large
part of the emission is relatively diffuse. We note that these
observations did not include LA and only partly the KP telescopes,
leading to a substantial loss of the short baselines required to
detect OH at angular scales $>80$ mas ($>325$ pc).

\begin{figure}[t]
  \epsscale{1} 
\plotone{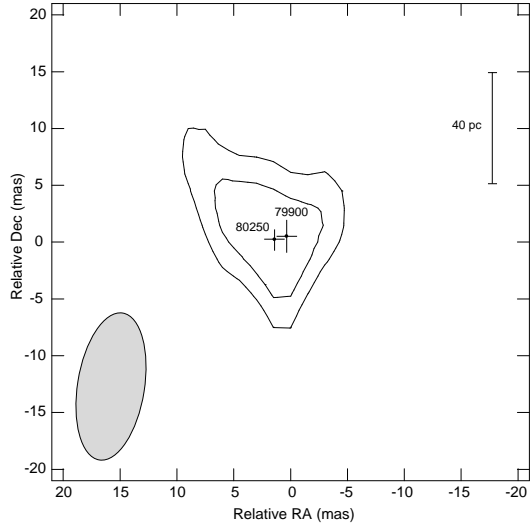}
\caption{The OH maser emission distribution in IRAS\,14070+0525, averaged over all channels with line emission. Contour levels are plotted starting at a 3$\sigma$ level of 1 mJy/beam and then increasing by a factor of $\sqrt 2$. The beam size indicated in the lower left corner is $13.01\times 5.87$ mas. The two crosses mark the positions of the two peaks identified in the VLBA spectrum, shown with 1$\sigma$ error bars.\label{ir14_average}}
\end{figure}

Figure \ref{ir14_average} displays the spatial extent of all emission
detected, which is contained within a region of around 10$\times$10
mas ($\sim 40\times 40$ pc). The emission is centered at RA 14 09
31.2446 and Dec 05 11 31.507 (J2000), which coincides with the optical
and NIR nuclear positions reported by \citet{kim02}.

\citet{condon98} measured a 1.4~GHz continuum flux density of 5 mJy.
No continuum emission was detected for IRAS\,14070+0525 at a 3$\sigma$
level of 0.45 mJy/beam.


\section{Discussion}

\subsection{IRAS\,12032+1707 Continuum}\label{contdisc}
The continuum in IRAS\,12032+1707 is only tentatively detected at the
shortest u,v distances (Sect.\ \ref{cont}). In part, this result could be
due to calibration errors, but it also indicates that the majority
of continuum emission is found on scales larger than 260 pc
(Sect.\ \ref{cont}). That implies an upper limit of the
brightness temperature to be around 3$\times 10^6$ K. Such a
brightness temperature and spatial distribution would be consistent
with a starburst origin similar to that seen in Arp\,220
\citep{smith98} and III\,Zw\,35 \citep{pihlstrom01}. 

Many ULIRGs fall on the radio-FIR correlation for starbursts
\citep{helou85, yun01}. This correlation can be understood by assuming
that the radio emission is arising from starburst induced supernova
remnants and from HII regions. Following the definitions from
\citet{helou85}, the correlation is quantified by a logarithmic
ratio between the far-infrared flux density ($FIR$) and the 1.4 GHz
radio flux density ($S_{\rm 1.4 GHz}$):

\begin{equation}
q={\rm log}\left(\frac{FIR}{3.75\times 10^{12} \rm W m^{-2}}\right)-{\rm log}\left(\frac{S_{\rm 1.4GHz}}{\rm W m^{-2}Hz^{-1}}\right)
\end{equation}

where $FIR$ is defined by

\begin{equation}
FIR=1.26\times10^{-14}(2.58S_{\rm 60 \mu m}+S_{\rm 100 \mu m}) \rm W m^{-2}
\end{equation}

For the infrared selected galaxies in the $IRAS$ 2 Jy sample, the mean
value of $q$ is $\simeq2.35$ \citep{yun01}. Given the FIR luminosity
and radio flux density of IRAS\,12032+1707, we estimate $q=1.77$. This
value appears low compared with the mean of 2.35 and could imply the
presence of an AGN. This appears contradictory to the non-detection of
a compact radio AGN core. However, radio-excess objects in the $IRAS$
2 Jy sample are defined as those objects having a radio luminosity
that is greater than 5 times larger that predicted by the radio-FIR
correlation.  Equivalently, this means objects for which q$\leq$1.64.
Furthermore, IRAS\,12032+1707 has a very large infrared luminosity
(log$(L_{\rm IR}/L_{\odot})=12.57$), placing the galaxy at the high
tail of the ULIRG luminosity distribution \citep{kim98}. For the most
luminous sources in the $IRAS$ 2 Jy sample, scattering from the
radio-FIR correlation is significantly larger than for the weaker
sources with weaker FIR flux densities.  Therefore, the relatively low
$q$-value for our source is not sufficient to indicate the presence of
an AGN component. Other evidence that IRAS\,12032+1707 is
starburst-powered comes from its relatively cool IR color, $F_{\rm
  25\mu m}/F_{\rm 60\mu m}=0.18$ \citep{kim98}. It is only for warm
colors ($F_{\rm 25\mu m}/F_{\rm 60\mu m}>0.25$) that an AGN component
is required to explain the resulting dust temperatures.

\subsection{IRAS\,12032+1707 OH and HI}\label{discuss}

The zero-intensity line width of the OH maser emission in
IRAS\,12032+1707 is around 2000 \kms, which makes this source the host
of one of the broadest OH megamaser lines. As mentioned in the
introduction, this velocity range could be due to a combination of
effects such as disk rotation and gas flows.  Due to the limited SNR
in our data, it is difficult to determine unambiguously the cause of
the masers in IRAS\,12032+1707. Below we discuss a few possible
scenarios that could agree with our observations.

\subsubsection{Multiple disks}\label{multipledisks}

It has been suggested that very broad OH megamaser lines might be a
combination of maser emission occurring in {\it both} of the merging
nuclei, as is the case for the OH megamaser prototype Arp\,220
\citep{diamond89}. We can rule out that maser emission in
IRAS\,12032+1707 is the combined emission from the southern and
northern nuclei, since it is clear that all maser emission occurs only
in the northern nucleus (Sect.\ \ref{ohspatial}).

We can not however exclude the possibility that the northern nucleus
itself is made up of two more closely interacting galaxies. In
Arp\,220 the nuclei have a projected separation of only 300 pc, which
would correspond to an angular resolution of 86 mas at the distance of
IRAS\,12032+1707, much smaller than the near-infrared seeing limit of
0.73\arcsec reported for the NIR image \citep{kim02}. This issue
requires further investigation by future higher resolution infrared
and optical imaging, combined with VLBI maser data.

In the OH megamaser source Mrk\,231 \hi\ absorption has been seen
against a faint continuum disk \citep{carilli98}, and the \hi\,
velocity field displays a gradient in the same direction as the OH
megamaser disk \citep{kloeckner03}. In particular, the \hi\ absorption
velocities encompass those of the OH lines. Assuming that \hi\ 
absorption in IRAS\,12032+1707 has the same close correlation to 
OH as the \hi\ in Mrk\,231, then the fact that \hi\ only partly
covers the OH velocities implies that the \hi\, absorption is
seen against only one of the disks.

\subsubsection{A single nuclear disk}\label{onedisk}

\begin{figure}[t]
  \epsscale{0.9}
  \rotatebox{-90}{
\plotone{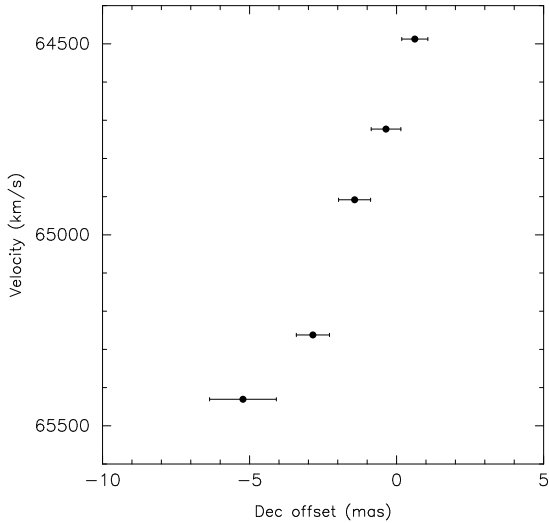}}
\caption{IRAS\,12032+1707: Velocity plotted as a function of position along the axis delineated by the three redshifted components.\label{velpos}}
\end{figure}

A clue to the nature of the OH maser emission must be given by the
plot of the centroid position of the different velocity components.
Clearly, there is a structured velocity field, with a bulk gradient in
the north-south direction. If we assume that all maser emission
belongs to a single disk-like structure, with a major axis going
through the three most redshifted points at a position angle (PA) of
11$^{\circ}$ (Fig.\ \ref{ir12_radecpos}), we can plot the data as
velocity versus distance along this axis (Fig.\ \ref{velpos}). A least
squares fit to the data points yields a gradient of 55 km\,s$^{\rm
  -1}$pc$^{\rm -1}$ over 17 pc. In total, this would mean an enclosed
mass of 2.2$\times 10^8$sin$^{\rm -2}i M_{\sun}$ within a radius of
8.5 pc. Such a velocity gradient is surprisingly high given the
gradients observed in other OH megamaser galaxies. For instance,
for both III\,Zw\,35 and Mrk\,231, the velocity gradient is close to 1.5
km\,s$^{\rm -1}$pc$^{\rm -1}$ \citep{pihlstrom01,kloeckner03}.

In contrast to the case of multiple disks suggested in Sect.\ 
\ref{multipledisks}, a scenario with a single disk for
IRAS\,12032+1707 makes it difficult to understand why the \hi\,
absorption covers only part of the OH velocity range (assuming that the
\hi\ and OH are parts of the same neutral gas structure).

\subsubsection{A disk and an inflow}

An alternative interpretation of Fig.\ \ref{ir12_radecpos} could
include two different gas components. A couple of facts support the
possibility of two physically different OH maser components in
IRAS\,12032+1707. Firstly, single-dish variability studies show that
the two strongest and most blueshifted peaks have a variable flux
density while the redshifted emission displays no significant
variability (J.\ Darling, priv.\ comm). Secondly, the \hi\ 
absorption velocities cover only the redshifted OH components (Sect.\ 
\ref{HIres}). If the \hi\ has a common origin with the redshifted OH
gas, the blueshifted and redshifted masers are likely to be physically
different gas structures.

Studying the relative location of maser components in Fig.\ 
\ref{ir12_radecpos}, we suggest that the two most blueshifted masers
(which also are the brightest ones) could represent the tangent points
of a circumnuclear disk.  With an inclined disk (or torus), the
tangent points will have the longest paths of velocity coherent gas,
resulting in regions of very strong maser emission.  This effect
has been observed in III\,Zw\,35 \citep{pihlstrom01}. The rotational
velocity of such a torus or disk structure in IRAS\,12032+1707 would
be 118 sin$^{-2}i$ \kms\, at a radius of 3 pc, yielding a modest
enclosed mass of 5$\times 10^6$sin$^{-2}i M_{\sun}$.

If the two blueshifted masers define the major axis of a disk
component, the redshifted masers are directed more or less
perpendicular to the disk (Fig.\ \ref{ir12_radecpos}). By studying the
optical and NIR images of IRAS\,12032+1707 \citep{veilleux02} it can
be seen that the southern optical nucleus is located slightly west of
south, with respect to the northern nucleus. Tentatively, the
redshifted masers could arise in gas that has been disturbed by tidal
interactions between the merging nuclei. With the caveat of
over-interpreting our low SNR data, it is tantalizing to note that
there is a weak hint in Fig.\ \ref{velpos} of gas closer to the
nucleus moving with higher velocity with respect to the nucleus, than
does the gas further away.

In this scenario we suggest that the \hi\ and the redshifted OH
originate in the same gas component. Any gas associated with the \hi\ 
absorption must thus necessarily be in front of the background
continuum. Assuming the blue masers define a disk, it is plausible
that the \hi\ is seen in absorption against a weak radio continuum
emitted by this disk. Given a limited field of view in the VLBA
observations, in addition to an only tentative detection of the radio
continuum, it is justified to question whether the radio continuum in
fact coincides at all with the maser emission. We note that
VLA A-array observations have showed that all radio continuum emission
originates in the northern optical nucleus (J.\ Darling, priv.\ 
comm.), and the OH maser emission is located in the same region. Since
\hi\ absorption requires the presence of a background continuum,
this is consistent with the \hi\ absorption and OH masers occurring in
the same region, at least on scales of a few kpc (corresponding to the
VLA A-array resolution).

The optical systemic velocity of the nucleus is 65055 \kms\, with an
1$\sigma$ error of $\pm70$\kms\,\citep{kim98}. Comparing the optical
systemic velocity with the OH and \hi\ velocities is hard since the
optical redshift is measured for the whole IRAS\,12032+1707 system,
and thus does not separate the optical nuclei. In addition, the
velocity is determined from optical emission lines that are often
broad and might not properly reflect the systemic velocity since the
emission lines may be associated with gas in motion with respect to
the nucleus. Nevertheless, within a 3$\sigma$ limit the systemic
velocity is consistent with redshifted maser components {\it
  falling in} towards the nucleus. \citet{veilleux99} reports on
optical emission lines in IRAS\,12032+1707, that are broad compared to
the typical ULIRG, consistent with the presence of an infalling gas
component.

Neutral gas flows have been observed previously in other galaxies. For
instance, VLA observations of the 1667, 1665 and 1612 MHz OH lines in
the Galactic Center have revealed gas streaming inwards from the CND
(the CircumNuclear Disk) toward SgrA* \citep{karlsson03}. The streamer can
be seen over a velocity range of $\sim$100 \kms.  Furthermore, \hi\ 
studies have extensively been used to trace merger dynamics, and one
example is the detailed studies of \hi\ emission in `The Antennae'.
\citet{hibbard01} have mapped high \hi\ column densities exceeding
$10^{22}$ cm$^{-2}$ in the region of the merging disks in this
source. It is not unlikely that similar column densities could be
probed by \hi\ absorption in a merger like IRAS\,12032+1707.


\subsection{IRAS\,14070+0525}

A major difficulty with interpreting the OH megamaser emission in
IRAS\,14070+0525 is that we only detect a fraction of the total single
dish emission. As already mentioned, a reason why so much emission
is resolved out may be the lack of short baselines in our
VLBA observations. The offset between the two detected peaks
is not significant. Optical and NIR imaging cannot distinguish more
than one nucleus in IRAS\,14070+0525, and this source is interpreted
as a merger where the nuclei have more or less coalesced
\citep{veilleux02}. Two nuclei could thus be located close to each
other, and violent gas kinematics adding to the line width would be
anticipated. More sensitive VLBI observations probing the emission on
all spatial scales will be needed to gain a better understanding of the OH
megamaser emission in this galaxy.

IRAS\,14070+0525 is classified optically as a Seyfert 2 galaxy
\citep{kim98}. However, the color given by $F_{\rm 25\mu m}/F_{\rm
  60\mu m}$ flux density ratio is only 0.13, and does not indicate the
presence of an AGN. Contrary to IRAS\,12032+1707, IRAS\,14070+0525
falls on the radio-IR correlation with $q=2.4$ \citep{condon98,kim98},
and no compact radio emission associated with an AGN was detected.
Hence, the radio emission is diffuse and consistent with a starburst.
Given the low 1.4 GHz flux density of 5\,mJy \citep{condon98}, a
slightly resolved starburst would be expected to fall below our
detection limit in these observations.



\section{Summarizing remarks}\label{conclusions}

We have presented VLBA data on the broad OH megamaser emission in 
two IRAS galaxies IRAS\,12032+1707 and IRAS\,14070+0525. Due to
insufficient coverage at short uv-spacings, only a minor part of the
OH flux density was detected in the second source, IRAS\,14070+0525.
We find no significant evidence of an ordered velocity field in this
source, but these results are inconclusive due to a very low SNR.

Almost all OH emission previously detected by single dish has been
recovered in IRAS\,12032+1707, and is found on a compact scale of
$<$100 pc. The emission shows an ordered velocity field that could be
consistent with a single disk. Although this may be the simplest
explanation, we prefer an alternative explanation that better ties
with the \hi\ absorption data and variability studies of the maser
emission. In this scenario, the two strongest and most blueshifted
maser features would be the tangent points of a disk with major axis in
the north-west to south-east direction. The remaining, redshifted
maser features are aligned roughly perpendicular to this disk,
extending to the south,  i.e., the direction in which the second 
optical nucleus
is located.  This redshifted emission could thus be associated with
tidally streaming gas falling onto the northern optical nucleus. The
\hi\ absorption covers the velocities of the redshifted maser
emission, suggesting a common origin.

We note that this scenario is speculative given the number of data points.
This should be confirmed with VLBI using a number of large dishes to
increase the SNR.  It would also be interesting to study high
resolution CO data of this source. CO is known to be a good tracer of
gas flows and disk structures in ULIRGs, and would thus give valuable
information about the molecular gas dynamics.

\section{Acknowledgments} 
The National Radio Astronomy Observatory is a facility of the National
Science Foundation operated under cooperative agreement by Associated
Universities, Inc. YMP wishes to acknowledge L.O.\ Sjouwerman, A.J.\ 
Mioduszewski and E.M.L.\ Humphreys for helpful comments on both the data
reduction as well as on the interpretation.


\end{document}